\documentclass[a4paper,12pt]{article}
\usepackage{graphicx}
\usepackage{psfrag}
\begin{document}
\title{The R\'enyi entropy $H_2$ as a rigorous, measurable lower
bound for the entropy of the interaction region in multiparticle production
processes }
\author{A.Bialas W.Czyz and K.Zalewski\thanks{Partly supported by the MEiN
grant 1 P03B 04529 (2005-2008).}
\\ M.Smoluchowski Institute of Physics
\\ Jagellonian University, Cracow\footnote{Address: Reymonta 4, 30 059 Krakow,
Poland, e-mail: bialas@th.if.uj.edu.pl, zalewski@th.if.uj.edu.pl}
\\ and\\ Institute of Nuclear Physics, Cracow}

\maketitle
\begin{abstract}
A model-independent lower bound on the entropy $S$ of the multi-particle system
produced in high energy collisions, provided by the measurable R\'enyi entropy
$H_2$, is shown to be very effective. Estimates show that the ratio $H_2/S$
remains close to one half for all realistic values of the parameters.
\end{abstract}

\section{Introduction}

The evaluation of the entropy of the stuff in the interaction regions of high
energy multiparticle production processes is an important problem (cf. e.g.
\cite{PAP},\cite{AKS2} and references given there). Within models, it is
possible to estimate this entropy with an uncertainty of about 10\% (cf. e.g.
\cite{PAP}). The problem we are addressing in the present paper is, how useful
is the model independent lower bound, which can be obtained from experiment
using the R\'enyi entropies.

Even in classical physics it is not easy to measure entropy. While there are
simple gadgets to measure temperature, pressure and volume, there is no
entropy-meter. The best one can do is to measure changes of entropy. This is
reasonably simple for reversible processes in closed systems, when the changes
of entropy in a volume are entirely due to the flows of heat across the
boundary. In general, however, there are sources of entropy within the volume
and these are more difficult to monitor. Moreover, in thermodynamics entropy is
primarily defined for equilibrium states. One extends the definition to
so-called incomplete equilibria where the system is thought of as consisting of
subsystems, each approximately in equilibrium and interacting weakly with each
other. Then the total entropy is calculated as the sum of entropies of these
subsystems interpreted as systems in equilibrium. It is not quite clear,
however, how far from equilibrium one can go without loosing the physical sense
of entropy.

Boltzmann proposed the definition of entropy

\begin{equation}\label{entsha}
  S = - \sum_ip_i\log p_i,
\end{equation}
where the sum is over all the states of the system, or in more rigorous texts
over all the states with non-zero probabilities, and $p_i$ is the probability
of state $i$. The scale of temperature is chosen here so that the Boltzmann
constant equals one. Thus entropy is dimensionless. When the probabilities are
given by the canonical ensembles this formula yields the results known from
standard thermodynamics; including the third principle of thermodynamics, which
states that under some conditions $\lim_{T\rightarrow 0}S(T,\ldots) = 0$.

A reinterpretation of formula (\ref{entsha}) in terms of information theory was
given by Shannon. The result is that the formula can be applied to any
probability distribution. In the present paper we deal with the particles
produced in high energy collisions of heavy ions. It is not clear how far from
equilibrium their distribution is. The possible final states, however,
certainly have a probability distribution. Therefore, to be on the safe side,
we will refer to Shannon's entropy, keeping in mind, however, that if the final
state is described by the (grand)canonical distribution, this is just
Boltzmann's entropy, or equivalently the entropy known from standard
thermodynamics.

\section{Estimate of entropy in the interaction region}

In order to get a rough estimate of the entropy in the interaction region let
us use the following simple model: a perfect gas of identical particles, with
mass $m$ each, contained in volume $V$. The temperature $T$ and the chemical
potential $\mu$ are given. Thus, neither the energy nor the number of particles
is fixed. Maxwell-Boltzmann statistics is used with a factor $\frac{1}{n!}$ in
the contribution of each $n$-particle state to the grand partition function.
This is sometimes called the quasi-classical approximation. A standard
calculation yields formula (\ref{entrmb}) for the entropy $S(m,V,T,\mu)$. In
order to get a number, however, it is necessary to know the values of the four
arguments. For the mass of the particle we will take the pion mass. According
to an estimate of Pal and Pratt \cite{PAP}, about half the entropy of the
system is carried by pions. Here we consider only one kind of pions, thus we
estimate about one sixth of the total entropy in the interaction region. The
chemical potential is usually put equal zero. For $\mu \rightarrow m$ from
below, Einstein's condensation of the gas takes place and the quasi-classical
approximation breaks down. Quantitative estimates of the chemical potential,
cf. e.g. \cite{AKS2}, \cite{TOW}, \cite{AKS}, give positive values, but
sufficiently far from $m$ to make the quasi-classical approximation good within
a few per cent. Temperatures are strongly model-dependent. In models where all
the transverse momenta are due to thermal motion, they can exceed $200$MeV. In
models, where much of the transverse momentum is due to collective motion they
can drop to about $100$MeV. For a recent discussion of the temperatures in RHIC
experiments see \cite{BIY}. In the following sections we will consider the
region

\begin{equation}\label{limits}
  \frac{m}{2} < T < \infty;\qquad -T < \mu < T; \qquad \mu < m.
\end{equation}
Here, for comparison with \cite{PAP}, we will use $T = 125 MeV$ and $\mu =
60$MeV, which corresponds to $\frac{\mu}{T} = 0.48$.

 The estimate of the interaction volume is by far the most difficult part of
the problem. The transverse dimensions are usually estimated from the data on
Bose-Einstein correlations. There are reasons, however, why these estimates are
rather uncertain. Let us mention just two. The phases of the elements of the
single-particle density matrix in the momentum representation cannot be
determined from momentum measurements. Neglecting these phases one can
underestimate the transverse size by a large factor \cite{BIZ}. Guessing them
one can err in either direction. Another reason is that according to the
standard formula for the diagonal elements of the density matrix in the
coordinate representation:

\begin{equation}\label{}
  \tilde{\rho}(\textbf{x},\textbf{x}) = \int\!\!\frac{d^3qd^3K}{(2\pi)^3}\rho(\textbf{K},\textbf{q})e^{i\textbf{qx}},
\end{equation}
where $\textbf{q} = \textbf{p - p}'$ and $\textbf{K} = \frac{1}{2}(\textbf{p +
p}')$, the mean squares of the components of $\textbf{x}$, which characterize
the size of the interaction region, are
\begin{equation}\label{}
  \langle x_i^2 \rangle =
  - \int\!\!d^3K\;\left(\frac{\partial^2}{\partial
  q_i^2}\rho(\textbf{K},\textbf{q})\right)_{\textbf{q}=0}.
\end{equation}
The small $\textbf{q}$ region, however, is experimentally inaccessible. The
data have to be extrapolated from larger $\textbf{q}$ regions. It is known that
Gaussian extrapolations yield underestimates, but steeper extrapolations can
give anything, infinity included. Other measures of the size in ordinary space,
e.g. from the half-width of the small $q^2$ peak in momentum space, are
difficult to interpret. Actually, even if the region $q \approx 0$ were know
there would be problems of interpretation. What happens there is strongly
affected by the halo of pions produced far away from the decays of long-lived
resonances. This would have to be somehow corrected.

The situation for the longitudinal dimension deserves an additional comment. At
high energies the total longitudinal size of the interaction region is much
larger than the transverse size. In fact, it is believed to increase linearly
with $\sqrt{s}$. The volume relevant for the calculation of entropy, however,
is the volume at given momentum. In the longitudinal direction this should
roughly correspond to the longitudinal size of the homogeneity region, which is
obtained from the study of the Bose-Einstein correlations with reservations as
for the transverse dimensions. The usual strategy is to calculate entropy
densities, or other ratios where the volume cancels. For instance, using
formula (\ref{entrmb}) and the corresponding formula for the particle
multiplicity we get

\begin{equation}\label{}
  \frac{S}{N} = 4 - \frac{\mu}{T}
  +\frac{m}{T}\frac{K_1\left(\frac{m}{T}\right)}{K_2\left(\frac{m}{T}\right)}.
\end{equation}
The number four on the right-hand side dominates. Each of the other two is of
the order of $0.4$ and, moreover, they tend to cancel. In particular, for $\mu
= 60$MeV the cancellation is almost exact and\footnote{Similar formulae have
been, of course, derived and used for many years.}

\begin{equation}\label{thumbr}
  \frac{dS}{dy} \approx 4 \frac{dN}{dy}.
\end{equation}
One could argue that in order to get results at fixed $y$ one should use a
two-dimensional momentum distribution. We checked that putting $p_L \equiv 0$
one has to replace the coefficient $4$ in this formula by $3.11$. Since,
however, for the experimental determination of $dN/dy$ one uses particles from
a rather large range of $p_L$, we consider the formula in the text more
realistic. For the $5$\% most central $Au-Au$ collisions at $\sqrt{s} =
130$GeV, for the $\pi^-$ mesons at midrapidity, the PHENIX collaboration finds
$\left(\frac{dN}{dy}\right)_{y=0} = 270 \pm 3$ \cite{ADC}. According to
\cite{PAP} this number should be reduced by 12\% in order to eliminate the
pions from decays of long living resonances. Thus we find

\begin{equation}\label{}
  \left(\frac{dS}{dy}\right)_{y=0} = 961.
\end{equation}
Pal and Pratt \cite{PAP}, from a more sophisticated, but closely related,
argument find for a sample of less central collisions (centrality $11$\%)
$\left(\frac{dS}{dy}\right)_{y=0} = 680$. For the corresponding sample
(centralities from 5\% to 15 \%) the PHENIX collaboration gives $\frac{dN}{dy}
= 200 \pm 2$, which introduced into (\ref{thumbr}) yields
$\left(\frac{dS}{dy}\right)_{y=0} = 704$ in reasonable agreement with the
number from \cite{PAP}. Similarly, using as the density of thermal $\pi^-$
mesons one half of the overall $\pi^-$ density, as recommended by Akkelin and
Sinyukov \cite{AKS2}, one obtains from (\ref{thumbr}), at $\sqrt{s} = 130$GeV
and $\sqrt{s} = 200$GeV respectively, $\frac{dS}{dy} = 520$ and $\frac{dS}{dy}
=610$ to be compared with 470 and 570 obtained in \cite{AKS2}. The particle
densities have been taken from the figures in ref. \cite{AKS2} and averaged
over the PHENIX and STAR results.

 Let us discuss now the
uncertainties of these numbers. The errors quoted here from \cite{ADC} are
statistical. Moreover, there is a systematic error estimated as 13\%
\cite{ADC}. In \cite{PAP} the error is estimated to be about 10\%. This error
would be quite acceptable, it is model dependent, however . For instance, most
crystals at low temperatures have very low entropies (third principle of
thermodynamics) and estimating their entropies from a perfect gas model would
give results wrong by orders of magnitude. Admittedly, the stuff in the
interaction region is not as regular as a crystal at low temperatures, but it
is not a perfect gas either. Few would doubt that the perfect gas approximation
is better than the low temperature crystal approximation, but how good it
actually is, is an open problem. In order to shed some light on this problem it
has been proposed by two of us \cite{BIC} to study the R\'enyi entropies, which
are both measurable and related to Shannon's entropy. For some more work in
this direction see \cite{BIC2}, \cite{BIC3} and for a recent review see
\cite{BIZ2}. Let us now recall some properties of R\'enyi entropies.

\section{R\'enyi entropies}

Let us consider an arbitrary system with states labelled by index $i$. We will
explicitly write down formulae for the case when $i$ is discrete, but the
generalization to the case when the spectrum of $i$ is continuous, or partly
continuous, can be done in the standard way.  The R\'enyi entropy of order $l$
is defined by the formulae

\begin{equation}\label{defren}
  H_l = \frac{\log C(l)}{1 - l}; \qquad C(l) = Tr \rho^l = \sum_i p_i^l,
\end{equation}
where $p_i$ is the probability of state $i$.

The R\'enyi entropies have a number of nice features (see e.g. \cite{BES}).
They are generalizations of Shannon's entropy, because $H_1 = S$. They satisfy
three out of the four axioms used by Khinchin to derive the formula for
Shannon's entropy, viz. they are functions of the probabilities $p_i$, they do
not change when a state with zero probability is added and they reach their
maximal values when all the probabilities $p_i$ are equal (micro-canonical
ensemble). It follows from (\ref{defren}) that for the micro-canonical ensemble
they coincide with Shannon's entropy, thus in a way they measure the
nonuniformity of the system. Instead of Khinchin's fourth axiom, which refers
to the entropy of a system consisting of subsystems, they all satisfy the
weaker condition that when a system consists of independent subsystems the
total entropy is the sum of the entropies of the subsystems.

A useful feature of Boltzmann's entropy, as studied in heavy ion collisons, is
that it changes neither during the free streaming of the particles after
freeze-out nor during the hydrodynamic expansion, provided this expansion is
non dissipative. R\'enyi entropies do not change in the free streaming process,
because they are invariant under the transformation corresponding to the
unitary time evolution of the system

\begin{figure}
\centering \psfrag{aaa}{\Large $\frac{m}{T}$} \psfrag{bbb}{\Large
$\frac{1}{N}(H_2 - S)$}
\includegraphics[totalheight = 4in]{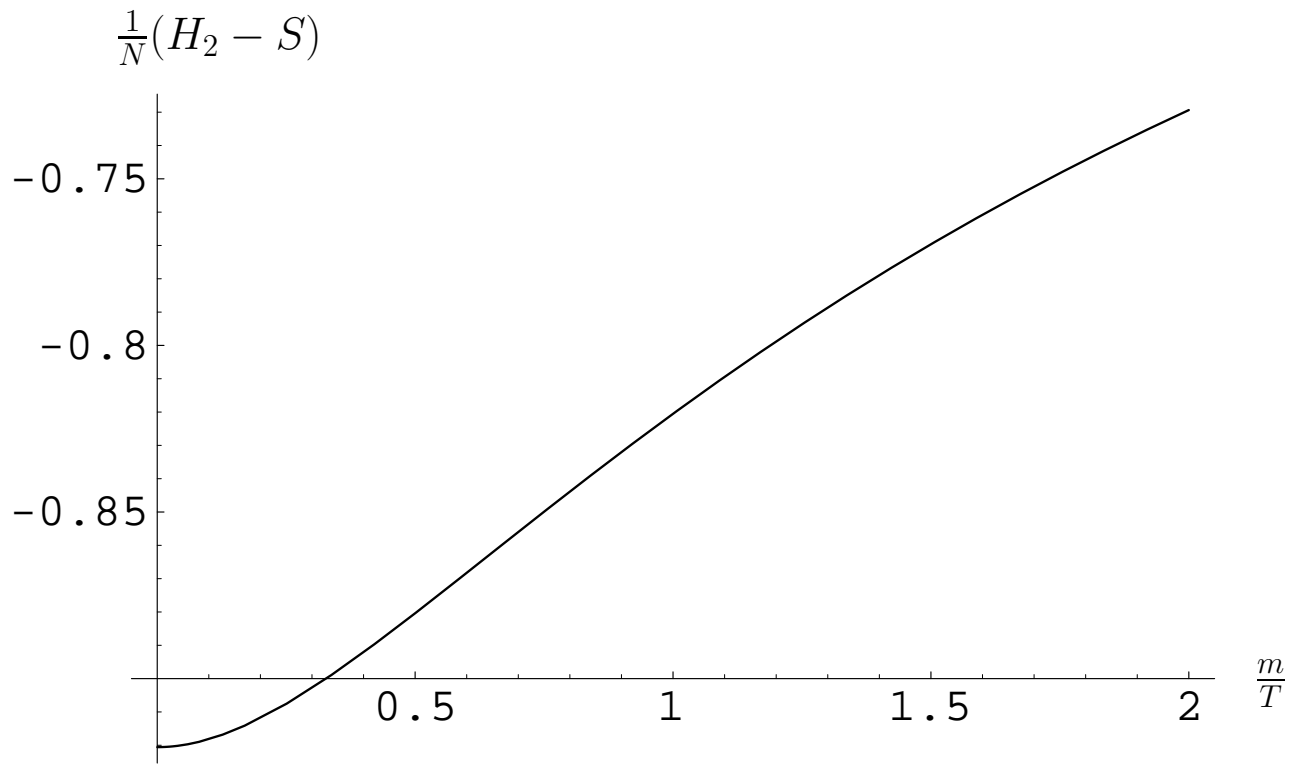}
\caption{Temperature dependence of the R\'enyi entropy $H_2$ along the
reversible adiabate, $S$ = Const, of a perfect gas. The number of particles $N$
is kept constant. The volume cancels.}
\end{figure}

\begin{equation}\label{}
  \rho(p,p') \rightarrow e^{-iHt}\rho(p,p')e^{iHt}.
\end{equation}
In fact, as seen from this argument, they are invariant even if the final state
interactions are included, provided these interactions can be described e.g. by
the Schr\"odinger equation.  They do change, however, when the volume of the
system is changed adiabatically and reversibly i.e. in particular in
non-dissipative flows. In order to illustrate this point we have calculated the
R\'enyi entropy $H_2$ for a perfect gas expanding adiabatically and reversibly,
i.e. at constant Shannon entropy. The result is shown in Fig. 1. Since
Shannon's entropy remains constant, the temperature dependence plotted reflects
the change of the R\'enyi entropy $H_2$ as a function of temperature. This
dependence is not very strong, however. For temperatures changing from very
high to $70$MeV the change is by about $0.2$ unit per particle, to be compared
with the total entropy of about four units per particle.

There are attempts to build a thermodynamics where the R\'enyi entropies play
the role of Boltzmann's entropy in standard thermodynamics (cf. e.g.
\cite{PAB}). The motivation is to describe distributions with "thick tails"
which occur often in experiment. Here we will not discuss these possibilities,
but use the R\'enyi entropies as a source of information about Shannon's
entropy. There is a theorem, cf. e.g. \cite{BES}, that the R\'enyi entropy
$H_l$ of a system is a decreasing function of its index $l$. A proof is given
in the Appendix. Therefore,

\begin{equation}\label{}
  S(m,V,T,\mu) > H_2(m,V,T,\mu).
\end{equation}
Similar inequalities hold also for the R\'enyi entropies of higher orders, but
the corresponding bounds are weaker and, therefore, less useful.

\section{R\'enyi entropies and standard thermodynamics}

Let us consider an open, one component system of particles in equilibrium, at
temperature $T$ and chemical potential $\mu$. The corresponding grand-canonical
probability distribution is

\begin{equation}\label{gracan}
  p_i = \frac{1}{\cal Z}e^{\frac{E_i - \mu n_i}{T}},
\end{equation}
where from the normalization of the probability distribution, the grand
partition function

\begin{equation}\label{}
  {\cal Z} = \sum_i e^{\frac{E_i - \mu n_i}{T}}.
\end{equation}
The grand partition function is related to the thermodynamic potential $\Omega$
by the formula

\begin{equation}\label{}
  \Omega(T,V,\mu) = -T\log {\cal Z}.
\end{equation}
Potential $\Omega$ is well known in statistical physics. Its relations to other
thermodynamical parameters follow from the identity

\begin{equation}\label{domega}
  d\Omega = -SdT - pdV - \sum_\alpha N_\alpha d\mu_{\alpha},
\end{equation}
where $\alpha$ labels the kinds of particles in the system. In the following we
discuss one component systems, so that the index $\alpha$ is not necessary and
we normalize the chemical potential $\mu$ so that $N$ denotes the number of
particles in the system.  If for arbitrary real, positive $\lambda$

\begin{equation}\label{}
\Omega(T,\lambda V,\mu) = \lambda \Omega(T,V,\mu),
\end{equation}
as is usually the case when there are no long range interactions, then
differentiating both sides of this equation with respect to $\lambda$, putting
$\lambda = 1$ and using (\ref{domega}) one finds

\begin{equation}\label{}
\Omega(T,V,\mu) = -pV.
\end{equation}

The probability distribution (\ref{gracan}) is a standard concept in the
statistical physics of equilibrium states,  but also for non-equilibrium states
it is considered a reasonable first guess, because it corresponds to no
information about the system except for the kind and average number of
particles, their average energy and the (fixed) volume  \cite{BES}.

Substituting formula (\ref{gracan}) into (\ref{defren}) and using the
definition of the potential $\Omega$ one finds

\begin{equation}\label{hkterm}
  H_l(T,V,\mu) = \frac{l}{l-1}\frac{1}{T}\left(\Omega(\frac{T}{l},V,\mu) -
  \Omega(T,V,\mu)\right).
\end{equation}
An equivalent formula has been given in \cite{BIC2}. Thus, in general, the
R\'enyi entropy $H_l$ is expressed by the thermodynamic functions at two
different temperatures $T$ and $T/l$. The only exception is at $l=1$, when the
definition of $H_l$ yields an indefinite expression of the type $0/0$. For $l
\rightarrow 1$, however, one finds the well-known result

\begin{equation}\label{}
  \lim_{l\rightarrow 1} H_l(T,V,\mu) = -\frac{\partial \Omega(T,V,\mu)}{\partial T}
  = S(T,V,\mu).
\end{equation}

\section{Gas of non-interacting free particles}

\begin{figure}
\centering
\includegraphics[totalheight = 4in]{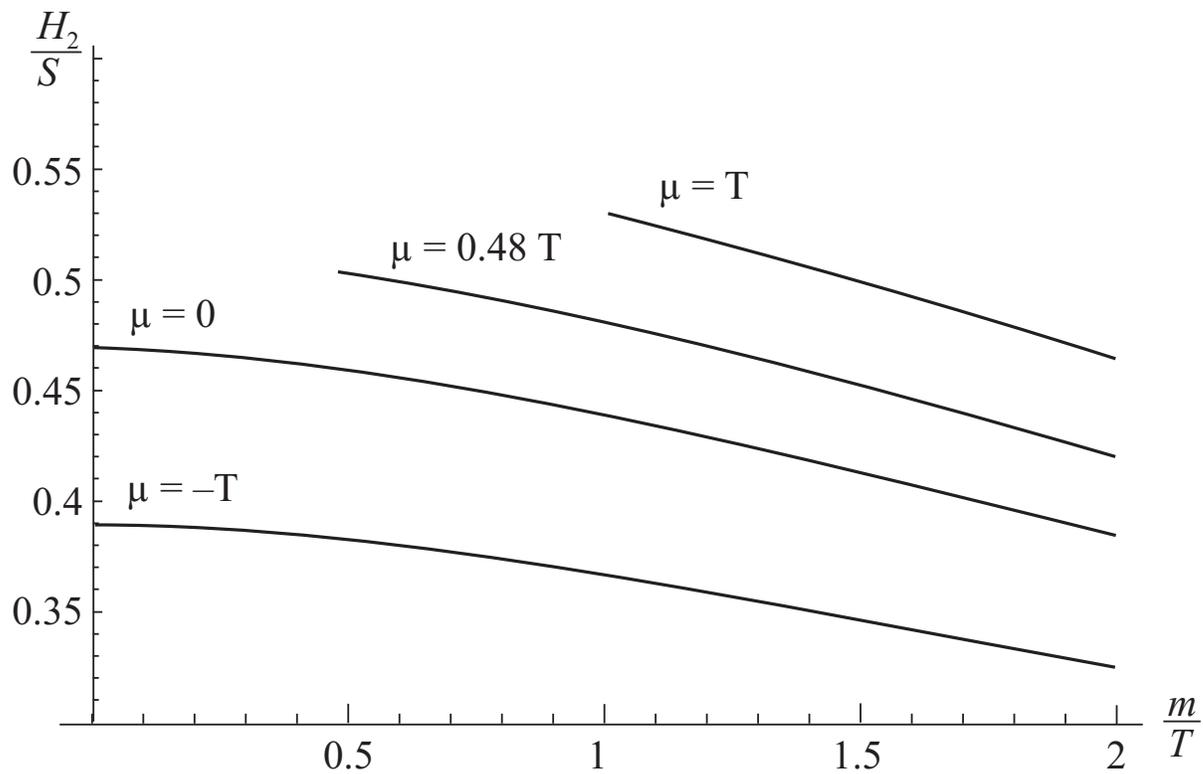}
\caption{Ratio of the R\'enyi entropy $H_2$ to Shannon's entropy $S$;
Maxwell-Boltzmann statistics. The unphysical results for $\mu > m$ are not
included}
\end{figure}

Let us consider a one component system of free spin zero particles, contained
in a fixed volume $V$ and with the Hamiltonian

\begin{equation}\label{hamilt}
  H = \sum_{i=1}^N \sqrt{p_i^2 + m^2}.
\end{equation}
The system is open so that the chemical potential $\mu$ and not the number of
particles $N$ is fixed. The state of the system $i$ is specified by giving the
number of particles $n_i$ in this state and all their momenta. Thus, in the
quasi-classical approximation, the grand partition function is

\begin{equation}\label{}
  {\cal Z}(T, V,\mu) = \sum_{n = 0}^\infty
  \frac{1}{n!}\left(\frac{V e^{\frac{\mu }{T}}}{(2\pi\hbar)^3}\int\!\!d^3p\;e^{-\frac{\sqrt{p^2 +
  m^2}}{T}}\right)^n
\end{equation}
Note that the indistinguishability of particles has been taken into account
only approximately, by dividing each $n$-particle contribution by $n!$. This is
the Maxwell-Boltzmann statistics. As is well known, it is a good  approximation
when the chemical potential is much smaller than the particle mass. A more
general formula based on the Bose-Einstein statistics will be derived latter.

The integral over momenta can be evaluated in spherical coordinates, using the
Bessel functions of imaginary argument $K_\nu(z)$ (cf. e.g. \cite{ABR}). The
resulting potential $\Omega(T,V,\mu)$ is

\begin{equation}\label{}
  \Omega(T,V,\mu) =
  -\frac{m^2VT^2}{2\pi^2\hbar^3}e^{\frac{\mu}{T}}K_2\left(\frac{m}{T}\right).
\end{equation}
The corresponding (Shannon) entropy is

\begin{equation}\label{entrmb}
  S = -\frac{\partial \Omega}{\partial T} =
  \frac{m^2VT}{2\pi^2\hbar^3}e^{\frac{\mu}{T}}\left[\left(4 - \frac{\mu}{T}\right)K_2\left(\frac{m}{T}\right) +
  \frac{m}{T}K_1\left(\frac{m}{T}\right)\right].
\end{equation}
Substituting the expression for $\Omega(T,V,\mu)$ into formula (\ref{hkterm})
one finds the R\'enyi entropies

\begin{equation}\label{hktose}
H_l(T,V,\mu) = \frac{l}{l-1}\frac{K_2\left(\frac{m}{T}\right) -
l^{-2}e^{\frac{(l-1)\mu}{T}}K_2\left(\frac{lm}{T}\right)}{\left(4 -
\frac{\mu}{T}\right)K_2\left(\frac{m}{T}\right) +
\frac{m}{T}K_1\left(\frac{m}{T}\right)}S(T,V,\mu).
\end{equation}
Let us consider some limiting cases. Using the formula \cite{ABR} valid for
$z\rightarrow 0$ and $\nu >0$

\begin{equation}\label{}
K_\nu(z) \approx \frac{1}{2}\Gamma(\nu)\left(\frac{2}{z}\right)^\nu,
\end{equation}
one finds that for $ m \ll T$ and $ km \ll T$

\begin{equation}\label{}
  H_l(T,V,\mu) \approx \frac{l}{l-1}\left(1 -
  l^{-4}e^{(l-1)\frac{\mu}{T}}\right)\frac{1}{4 - \frac{\mu}{T}}S(T,V,\mu).
\end{equation}
This formula is useful for numerical estimates, because in the limit $m
\rightarrow 0$ formula (\ref{hktose}) becomes an $\frac{\infty}{\infty}$
expression.

Making the further assumption $\mu \rightarrow 0$ one finds the formula given
in \cite{BIC}

\begin{equation}\label{}
  H_l(T,V,\mu) \approx \frac{1 + l^{-1} + l^{-2} + l^{-3}}{4 }S(T,V,\mu).
\end{equation}

The ratio $\frac{H_2(T,V,\mu)}{S(T,V,\mu)}$, calculated numerically from
formula (\ref{hktose}), is shown in Fig. 2. The values for $\mu > m$ are not
shown, because they are unphysical. One should keep in mind, however, that also
for $\mu < m$ our method of handling the indistinguishability of massive
particles is quantitatively reliable only for $\frac{m - \mu}{T}$ sufficiently
large.

The results are rather encouraging. The ratio decreases with increasing
$\frac{m}{T}$ and with decreasing $\frac{\mu}{T}$, but even for $\mu = -T$ and
$m = 2T$ it is still about $0.33$ so that the lower bound underestimates the
exact value by a factor of three. At $m=\mu=T$ the ratio reaches its maximum of
about $0.53$. This last result, however, should be verfied, because the
arguments are beyond the reach of applicability of the Maxwell-Boltzmann
statistics. Putting $p_L \equiv 0$ one finds qualitatively similar results with
the ratio $H_2/S$ ranging, in the $\{T,\mu \}$ region considered, from $0.35$
to $0.875$.

\begin{figure}
\centering
\includegraphics[totalheight = 4in]{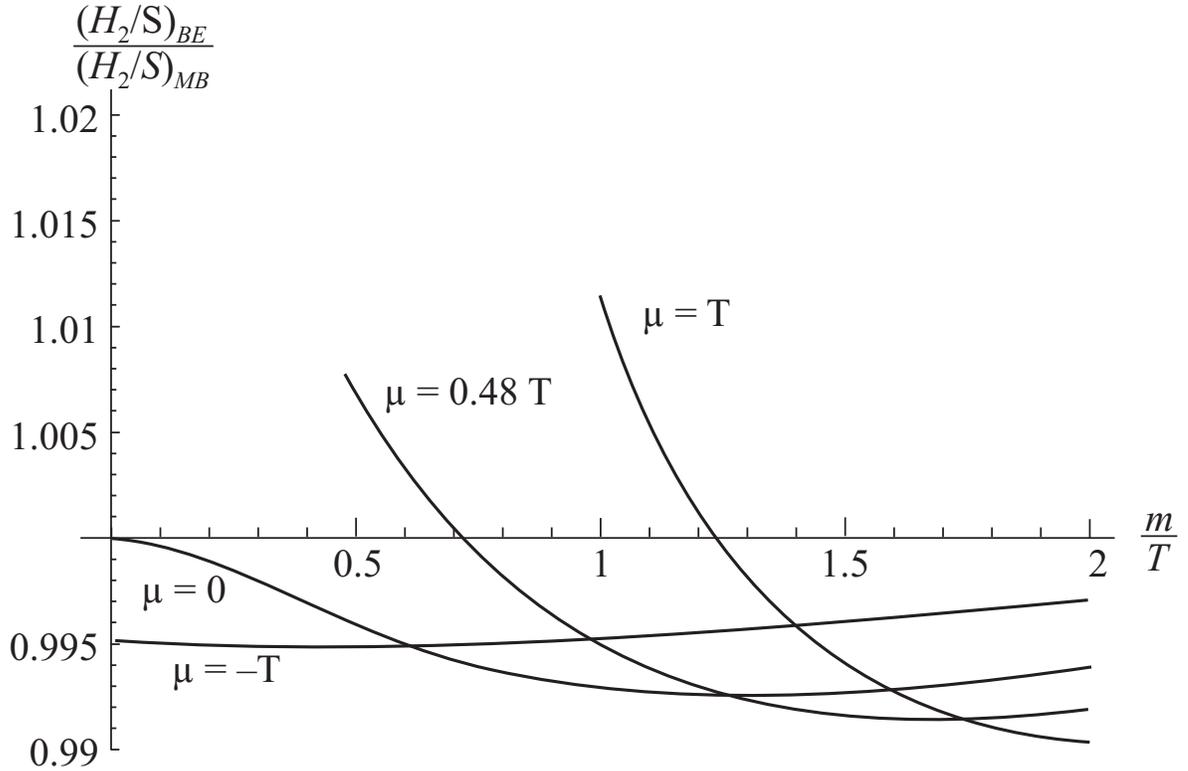}
\caption{Double ratio $\frac{(H_2/S)_{BE}}{(H_2/S)_{MB}}$ . The unphysical
results for $\mu > m$ are not included.}
\end{figure}

Let us reconsider now the gas of non-interacting free particles with the
Hamiltonian (\ref{hamilt}), but handling correctly the statistics. The standard
method is to use, instead of the states characterized by the number of
particles $n_i$ and momenta of the $n_i$ particles, the occupation numbers
$n_{\textbf{p}}$ of all the single particle states. Then

\begin{equation}\label{}
  {\cal Z}(T,V,\mu) = \prod_\textbf{p}\left(\sum_{n(\textbf{p})=0}^\infty
  e^{-n(\textbf{p})\frac{E(p) - \mu}{T}}\right)
\end{equation}
Summing the geometrical progressions and using the quasi-classical
approximation to convert in $\log{\cal Z}$ the summation over momenta into an
integration one finds

\begin{equation}\label{}
  \Omega(T,V,\mu) = \frac{VT}{(2\pi)^3} \int\!\!d^3p\;\log\left(1 -
  e^{-\frac{E(p)-\mu}{T}}\right).
\end{equation}
Expanding the logarithm in powers of the exponential and integrating term by
term like in the Maxwell-Boltzmann case one finds

\begin{equation}\label{}
  \Omega(T,V,\mu) = - \frac{m^2VT^2}{2\pi^2}\sum_{n=1}^\infty
  n^{-2}e^{\frac{n\mu}{T}}K_2\left(\frac{nm}{T}\right) .
\end{equation}
Note that keeping the $n=1$ term only, one reproduces the Maxwell-Boltzmann
case. The corresponding Shannon entropy is

\begin{equation}\label{}
  S(T,V,\mu) = \frac{m^2VT}{2\pi^2}\sum_{n=1}^\infty
  n^{-2}e^{\frac{n\mu}{T}}\left[\left(4 - \frac{n\mu}{T}\right)K_2\left(\frac{nm}{T}\right) +
  \frac{nm}{T}K_1\left(\frac{nm}{T}\right)\right] .
\end{equation}
Let us note the identities

\begin{equation}\label{}
  \Omega(T,V,\mu) = \sum_{n=1}^\infty \Omega_{MB}(\frac{T}{n},V,\mu);\qquad
  S(T,V,\mu) = \sum_{n=1}^\infty n^{-1}S_{MB}(\frac{T}{n},V,\mu),
\end{equation}
where the subscript $MB$ denotes the quantities calculated in the
Maxwell-Boltzmann approximation. The R\'enyi entropies are

\begin{equation}\label{}
H_l(T,V,\mu) = \frac{l}{l-1} \frac{\sum_{n=1}^\infty
(n)^{-2}e^{\frac{nm}{T}}K_2\left(\frac{nm}{T}\right) - \sum_{n=1}^\infty
(ln)^{-2}e^{\frac{lnm}{T}}K_2\left(\frac{lnm}{T}\right)} {\sum_{n=1}^\infty
n^{-2}e^{\frac{n\mu}{T}} \left[ \left(4 -
\frac{n\mu}{T}\right)K_2\left(\frac{nm}{T}\right) +
\frac{nm}{T}K_1\left(\frac{nm}{T}\right) \right]}S(T,V,\mu).
\end{equation}

In Fig. 3 the double ratio $\frac{(H_2/S)_{BE}}{(H_2/S)_{MB}}$ is shown. It is
seen that the ratio of the second R\'enyi entropy to Shannon's entropy changes
by less than about one per cent when statistics is changed from
Maxwell-Boltzmann to Bose-Einstein. On the other hand, as seen from Fig. 4, the
entropy itself changes by up to 16\%.

\begin{figure}
\centering
\includegraphics[totalheight = 4in]{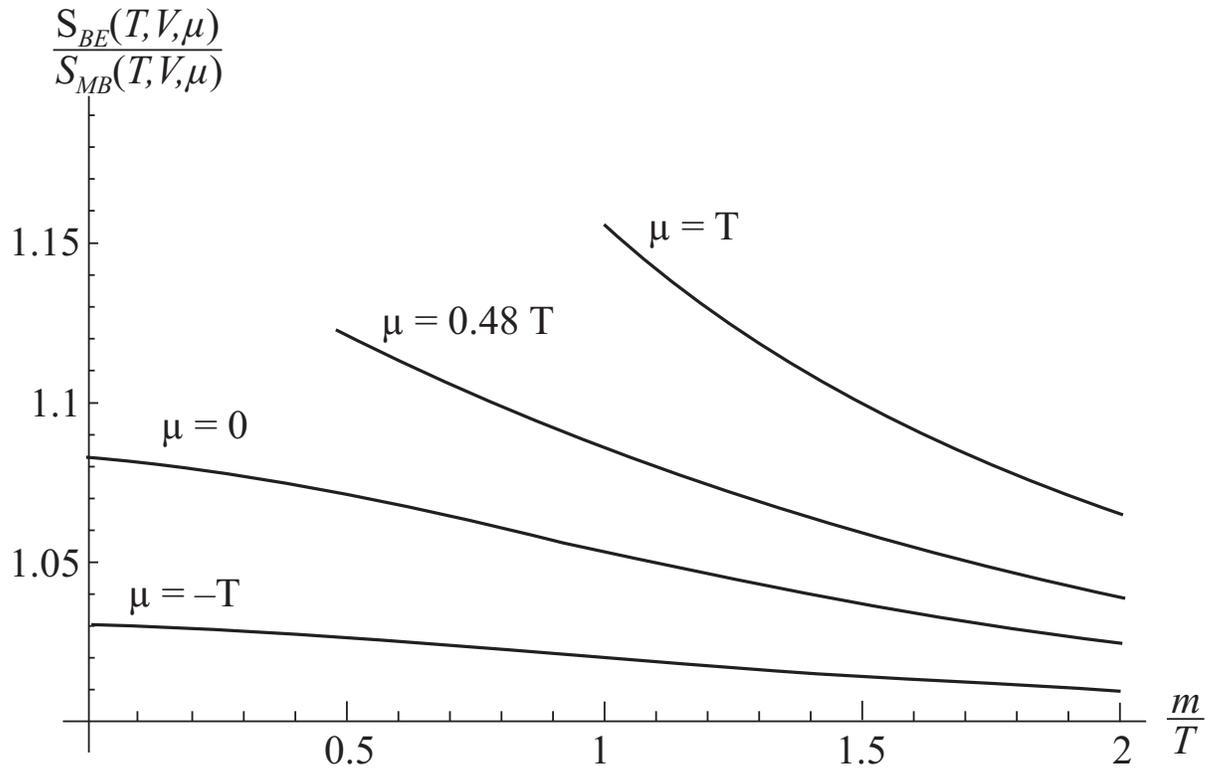}
\caption{Ratio of Shannon entropies: $\frac{S_{BE}(T,V,\mu)}{S_{MB}(T,V,\mu)}$
 The unphysical results for $\mu > m$ are not included.}
\end{figure}

\section{Gas of noninteracting particles in an external potential}

The gas studied in the preceding section was confined in a constant volume. It
may be more realistic to assume that the volume increases with increasing
energy per particle, i.e. with increasing temperature. A model of this type can
be obtained by replacing the free particle Hamiltonian (\ref{hamilt}) by a
Hamiltonian including an external potential. We will discuss the simple case of
the harmonic oscillator potential:

\begin{equation}\label{}
  H = \sum_{i=1}^N \left(\sqrt{p_i^2 + m^2} + \frac{1}{2}K\textbf{x}_i^2\right),
\end{equation}
where $K$ is a constant. Repeating the analysis from the preceding section one
finds that in the calculation of $\log{\cal Z}$ the only difference is that the
volume $V$ is replaced by

\begin{equation}\label{}
  V_{eff}(T) = \int\!\!d^3x\;e^{-\frac{K\textbf{x}^2}{2T}} = V_0
  T^{\frac{3}{2}};\qquad V_0 = (\frac{2\pi}{K})^{\frac{3}{2}}.
\end{equation}
 Thus

\begin{eqnarray}\label{}
\Omega(T,\mu) &=&
- \frac{m^2V_0}{2\pi^2\hbar^3}T^{\frac{7}{2}}K_2\left(\frac{m}{T}\right)e^{\frac{\mu}{T}},\\
S(T,\mu) &=&
\frac{m^2V_0}{2\pi^2\hbar^3}T^{\frac{5}{2}}\left[K_2\left(\frac{m}{T}\right)\left(\frac{11}{2}
- \frac{\mu}{T} \right) +   \frac{m}{T}K_1\left(\frac{m}{T}
\right)\right]e^{\frac{\mu}{T}}
\end{eqnarray}
and the R\'enyi entropies are

\begin{equation}\label{}
  H_l(T,\mu) = \frac{l}{l-1}
  \frac{
  K_2\left(\frac{m}{T}\right) -  l^{-\frac{7}{2}}
  K_2\left(\frac{lm}{T}\right)e^{(l-1)\frac{\mu}{T}}}
 {K_2\left(\frac{m}{T}\right)\left(\frac{11}{2} - \frac{\mu}{T}\right) +
 \frac{m}{T}K_1\left(\frac{m}{T}\right)}
  S(T,\mu).
\end{equation}
The ratio $\frac{H_2(T,\mu)}{S(T,\mu)}$ is plotted in Fig. 5. It is seen that
the presence of the potential reduces the ratio $\frac{H_2(T,\mu)}{S(T,\mu)}$.
In the region shown in the graph the ratio is between $0.26$ and $0.40$.
Qualitatively, the dependence on the parameters is as before.

\begin{figure}
\centering
\includegraphics[totalheight = 4in]{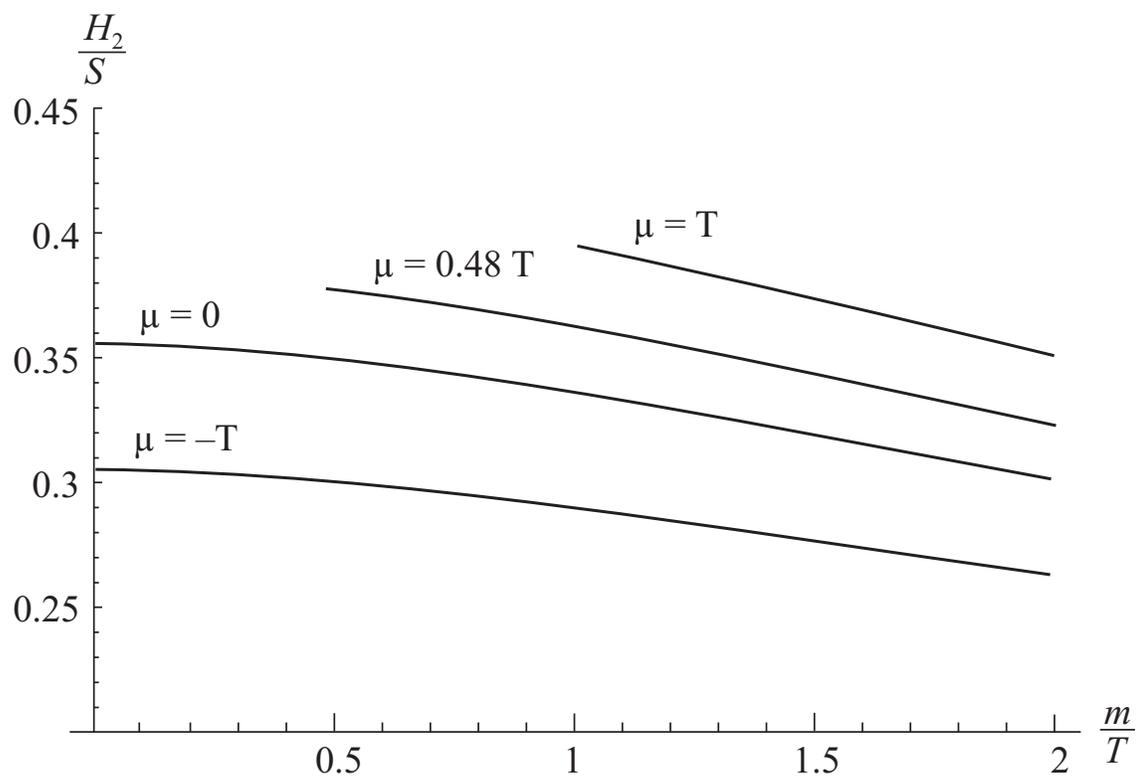}

\caption{Ratio $H_2/S$ for non-interacting particles in the harmonic oscillator
potential; quasi-classical approximation. The unphysical results for $\mu > m$
are not included}
\end{figure}

\section{Conclusions}

 The entropy $S$ is rather difficult to estimate directly from the
 data. Recently \cite{BIZ}, we have proposed a method to measure the R\'enyi
 entropy $H_2$, which provides a rigorous lower bound for $S$. In the present
 paper we investigate the relation between $H_2$ and $S$ in order to determine
 how close to the actual value of $S$ this bound is. Using the ideal gas model
 we find that, for the relevant (rather wide) range of parameters,
 $H_2$ is not far from $\frac{1}{2}S$. The detailed results are presented and
 discussed in the text. It is found that the ideal gas model reproduces
 within $10$\% the entropy densities obtained by other authors using more
 sophisticated methods \cite{PAP}, \cite{AKS2}. This suggests that also our
 estimate of the ratio $H_2/S$ should hold in more realistic models. We,
 therefore, conclude that if the measured R\'enyi entropy $H_2$ turns out to be
 much smaller than half the entropy $S$ estimated from a model, the model is unlikely to be realistic.

 The authors thank Ewa Gudowska-Nowak, Mariusz Sadzikowski and Karol \.{Z}yczkowski for
 discussions and Yuri Sinyukov for calling their attention to ref. \cite{AKS2}.

\section{Appendix}
The R\'enyi entropy $H_l$ is a decreasing function of the index $l$. This can
be seen as follows (cf. \cite{BES}). Differentiating both sides of the
definition (\ref{defren}) with respect to $l$ we get

\begin{equation}\label{}
  \frac{dH_l}{dl} = - (1-l)^2\sum_iP_i\log\frac{P_i}{p_i},
\end{equation}
where the  notation

\begin{equation}\label{}
  P_i = \frac{p_i^l}{\sum_ip_i^l}
\end{equation}
has been introduced. Using the identity

\begin{equation}\label{}
  \log x \geq 1 - \frac{1}{x},
\end{equation}
where equality holds only when $x=1$, one easily checks that for all $l \neq 1$
the right-hand side is non-positive. Actually, it is negative unless all the
probabilities $p_i$ are equal, which is not the case for multiple particle
production processes.

\end{document}